\documentclass[%
reprint,
amsmath,amssymb, 
aps,
prb
]{revtex4-2}
\usepackage{graphicx, color}
\usepackage{dcolumn}
\usepackage{bm}
\usepackage{appendix,float}
\usepackage[colorlinks,urlcolor=blue,linkcolor=blue,anchorcolor=blue,citecolor=blue]{hyperref} 

\begin{document}
\preprint{APS/123-QED}
\title{Crystal valley Hall effect}
\author{Chao-Yang Tan$^{1,2}$}
\author{Ze-Feng Gao$^{1,2}$}
\author{Huan-Cheng Yang$^{1,2}$}
\author{Zheng-Xin Liu$^{1,2}$}
\author{Kai Liu$^{1,2}$}
\author{Peng-Jie Guo$^{1,2}$}
\email{guopengjie@ruc.edu.cn}
\author{Zhong-Yi Lu$^{1,2,3}$}
\email{zlu@ruc.edu.cn}
\affiliation{1. Department of Physics and Beijing Key Laboratory of Opto-electronic Functional Materials $\&$ Micro-nano Devices. Renmin University of China, Beijing 100872, China}
\affiliation{2. Key Laboratory of Quantum State Construction and Manipulation (Ministry of Education), Renmin University of China, Beijing 100872, China}
\affiliation{3. Hefei National Laboratory, Hefei 230088, China}
\date{\today}
\begin{abstract}

Time-reversal symmetry is thought to be a necessary condition for realizing the valley Hall effect. However, if time-reversal symmetry is broken, whether the valley Hall effect can be realized has not been explored. In this paper, based on symmetry analysis and first-principles electronic structure calculations, we demonstrate that the valley Hall effect without time-reversal symmetry can be realized in two-dimensional altermagnetic materials Fe$_2$WSe$_4$ and Fe$_2$WS$_4$. Due to crystal symmetry being required, the valley Hall effect without time-reversal symmetry is termed the crystal valley Hall effect. In addition, under uniaxial strain, both monolayer Fe$_2$WSe$_4$ and Fe$_2$WS$_4$ can realize the piezomagnetic effect. Under biaxial compressive stress, both monolayer Fe$_2$WSe$_4$ and Fe$_2$WS$_4$ will transform from the altermagnetic semiconductor phase to the bipolarized topological Weyl semimetal phase. Our paper not only provides another direction for exploring the valley Hall effect but also provides a good platform for exploring altermagnetic semiconductors and altermagnetic topological phase transitions.

\end{abstract}

\maketitle

{\it Introduction.} 
Time-reversal ($T$) symmetry results in band degeneracy but opposite Berry curvatures at $k$ and $-k$ in momentum space. For two-dimensional materials, the Berry curvature can generate an anomalous velocity perpendicular to the direction of an in-plane electric field. If a two-dimensional nonmagnetic insulator without space-inversion symmetry has only two valleys connected by the $T$ symmetry, for example, $k$ and $-k$ valleys in monolayer MoS$_2$, the electrons from the two valleys with opposite Berry curvatures will move in opposite directions under the in-plane electric field. This is known as valley hall effect. The valley Hall effect has not only been theoretically proposed but also experimentally realized \cite{Graphene-2007, VHE-MoS2-2012, MoS2-Science2014}. Thus, the time-reversal symmetry is thought to be a necessary condition for realizing the valley Hall effect. 
Nevertheless, it is a fundamental issue whether the valley Hall effect can still be realized if the $T$ symmetry is broken.

Recently, altermagnetism as another magnetic phase has been proposed, which is distinguished from ferromagnetism and antiferromagnetism \cite{altermagnetism-1,altermagnetism-2,altermagnetism-3,altermagnetism-4,PRX-1,PRX-2,PRX-3,QAH-npj2023}. Altermagnetism has the duality of real-space antiferromagnetism and reciprocal-space anisotropic spin polarization similar to ferromagnetism. Moreover, altermagnetic materials will break $T$ symmetry under spin-orbital coupling (SOC). Thus, altermagnetic materials can realize many unique physical effects, such as spin-splitting torque \cite{SST-PRL2021,SST-NE2022,SST-PRL2022,SST-PRL2022-2}, giant magnetoresistance effect \cite{GMR-PRX2022,GMR-2024}, tunneling magnetoresistance effect \cite{GMR-PRX2022,TMR-Shao2021}, nontrivial superconductivity \cite{SC-AM}, time-reversal odd anomalous effect \cite{AHE-Sinova2022,AHE-RuO2-NE2022,AHE-MnTe-PRL2023,AHE-hou2023,altermagnetism-2,MOE-Yao2021,CTHE-Yao2024}, quantum anomalous Hall effect \cite{QAH-npj2023}, higher-order topological states \cite{HighoT-liu2024}, Majorana corner modes \cite{MCM-liu2023}, piezomagnetism effect \cite{piezomagnetism-NC}, altermagnetic ferroelectricity \cite{LiFe2F6-guo2023}, strong spin-orbit coupling effect in light element altermagnetic materials \cite{NiF3-qu2024}, and so on. Remarkably, altermagnetic materials can have valley electrons with opposite spin polarization deriving from crystal symmetry. This provides a possibility for realizing valley Hall effect without the $T$ symmetry.

In this paper, based on symmetry analysis and first-principles electronic structure calculations, we demonstrate that crystal valley Hall effect can be realized in two-dimensional altermagnetic Fe$_2$WSe$_4$ and Fe$_2$WS$_4$. Moreover, our calculations show that both monolayer Fe$_2$WSe$_4$ and Fe$_2$WS$_4$ have magnetic transition temperatures higher than room temperature, which are 330K and 500K, respectively. In addition, under uniaxial strain, both monolayer Fe$_2$WSe$_4$ and Fe$_2$WS$_4$ generate valley polarization. Meanwhile, under biaxial compressive stress, a topological phase transition from the altermagnetic semiconductor phase to bipolarized Weyl semimetal phase takes place in both monolayer Fe$_2$WSe$_4$ and Fe$_2$WS$_4$.


{\it Methods.} The electronic structure calculations were performed based on the VIENNA AB INITIO SIMULATION PACKAGE (VASP) \cite{Vasp-1996} with the projector augmented wave (PAW) method \cite{PAW-1994}. The Perdew-Burke-Ernzerhof (PBE) exchange-correlation functional within the generalized gradient approximation (GGA) was used in our calculations \cite{GGA-1996}. The energy cutoff 600 eV, energy convergence criterion $10^{-6}$ $\rm{eV}$, force convergence criteria $10^{-3}$ $\rm{eV}/\text{\r{A}}$, and k-mesh $12\times12\times1$ ($\Gamma$ centered Monkhorst-Pack) were used. The dynamical stability of Fe$_2$W$X_4$ was confirmed by using the density functional perturbation theory, adopting a $3\times3\times1$ supercell and $5\times5\times1$ k-meshes ($\Gamma$-centered Monkhorst-Pack). The maximally localized Wannier functions were constructed using the WANNIER90 package \cite{wan90-2008}. The Berry curvature was calculated using POSTW90 code. Finally, the Monte Carlo simulations based on the classical Heisenberg model were performed by using the MCSOLVER \cite{mcTN-2019}.

\begin{figure}[htbp]
\centering
\includegraphics[width=8.5cm]{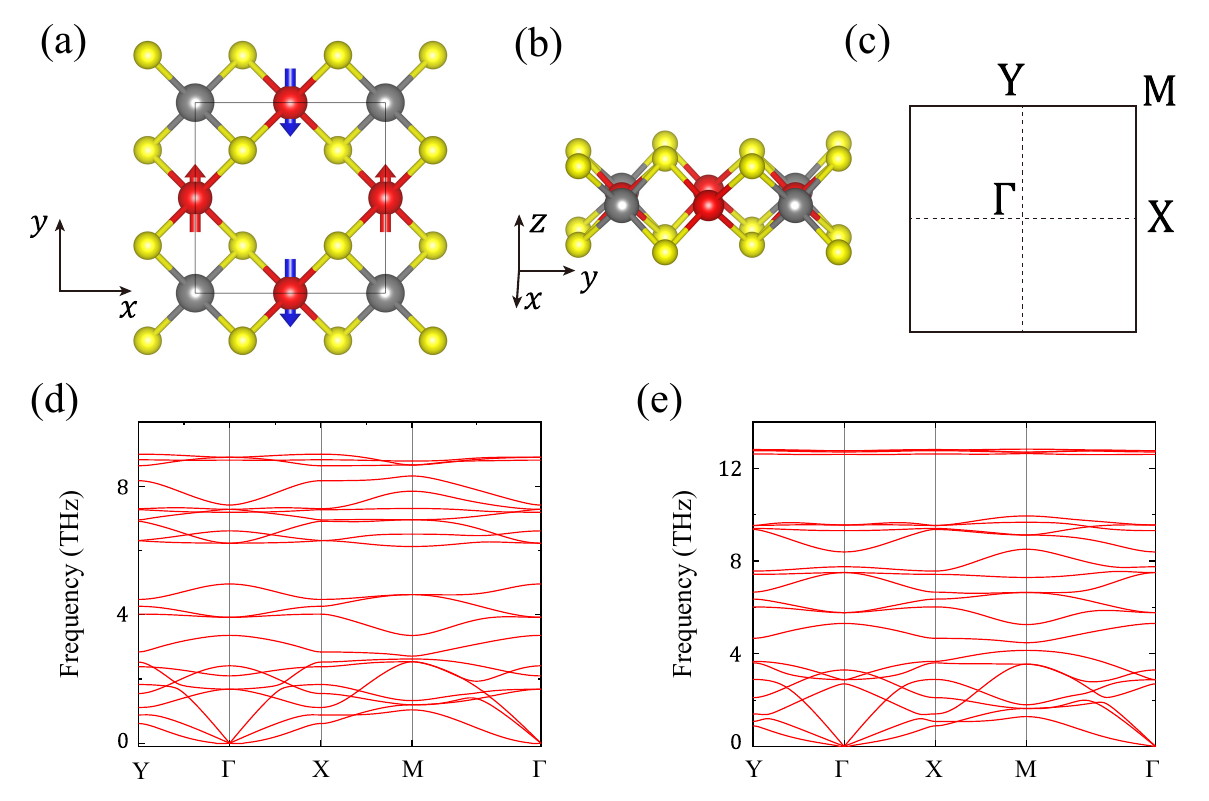}
\caption{(a), (b) Top and side views of crystal and magnetic structures of Fe$_2$W$X_4$, respectively. The red and blue arrows represent spin-up and spin-down magnetic moments, respectively. (c) The Brillouin zone (BZ) of Fe$_2$W$X_4$ where the high-symmetry points are labeled. The phonon dispersion for (d) Fe$_2$WSe$_4$ and (e) Fe$_2$WS$_4$.}
\label{fig1}
\end{figure}

{\it Results and analysis.} Monolayer Fe$_2$W$X_4$ ($X$=Se,S) takes a square lattice structure with the symmorphic space group $P$-$42m$ (No. 111) symmetry and the corresponding point group is $D_{2d}$ with generators $S_{4z}$ and $C_{2x}$. Monolayer Fe$_2$W$X_4$ contains three atomic layers where the Fe and W atomic layer is sandwiched by two X atomic layers as shown in Figs. \ref{fig1}{(a)} and \ref{fig1}{(b)}, and the corresponding BZ with high-symmetry points is shown in Fig.~\ref{fig1}{(c)}. The crystal parameters of monolayer Fe$_2$WX$_4$ are obtained by structural relaxation. The crystal parameters of monolayer Fe$_2$WSe$_4$ and Fe$_2$WS$_4$ are 5.572 $\text{\r{A}}$ and 5.436 $\text{\r{A}}$, respectively. Then we calculated the phonon spectrum of monolayer Fe$_2$W$X_4$, which are shown in Figs. \ref{fig1}{(d)} and \ref{fig1}{(e)}. From Figs. \ref{fig1}{(d)} and \ref{fig1}{(e)}, the phonon spectrum of both monolayer Fe$_2$WSe$_4$ and Fe$_2$WS$_4$ have no imaginary frequency. Therefore, both monolayer Fe$_2$WSe$_4$ and Fe$_2$WS$_4$ are dynamically stable.

Furthermore, the crystal structure of monolayer Fe$_2$W$X_4$ is proposed based on the synthesized layered Cu$_2$W$X_4$ ($X$=S,Se) \cite{Cu2MX4-1993,Cu2MX4-2005,Cu2MX4-2019,Cu2MX4-2022}, Cu$_2$MoS$_4$\cite{CuMoS-2012} and Ag$_2$WS$_4$ \cite{Ag2WS4-2018}, thus it is hopeful that monolayer Fe$_2$W$X_4$ may be synthesized experimentally. 

\begin{figure*}[htbp]
\centering
	\includegraphics[width=15cm]{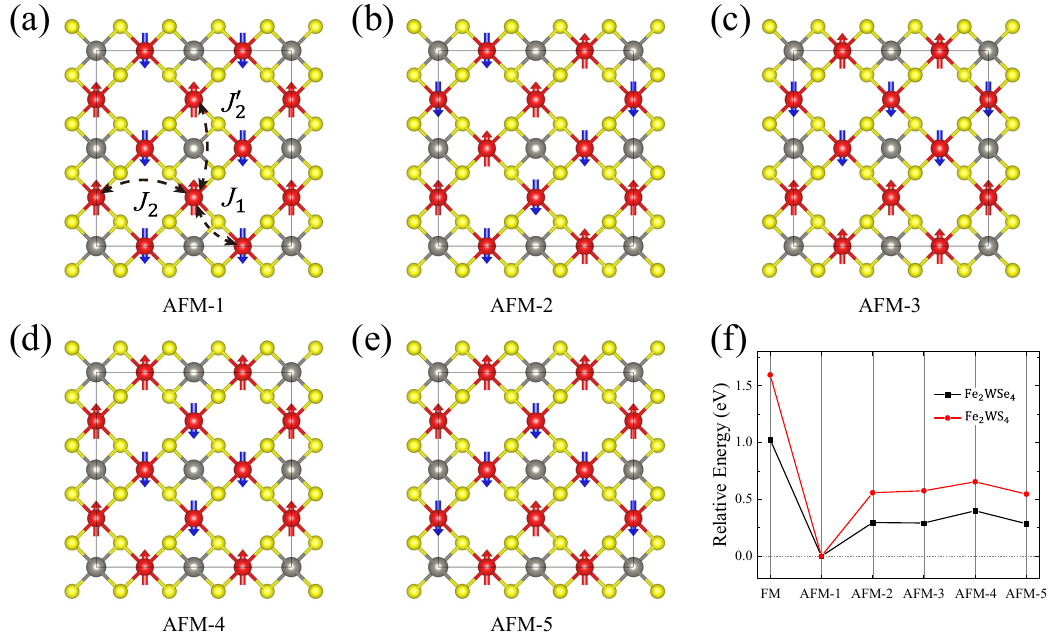}
	\caption{(a)-(e) Five different antiferromagnetic configurations with $2 \times 2\times 1$ supercell.  The red and blue arrows represent spin-up and spin-down magnetic moments, respectively. The indexes $J_1$, $J_2$ and $J_2^\prime$ in (a) denote the diagram of nearest and next-nearest exchange interaction parameters. (f) Relative energies of five magnetic states with respect to the AFM-1 state.}
	\label{fig2}
\end{figure*}

To determine the magnetic ground states of Fe$_2$W$X_4$, we consider six different magnetic configurations including one ferromagnetic (FM) and five antiferromagnetic as shown in Figs. \ref{fig2}{(a)}-\ref{fig2}{(e)} within a $2\times2\times1$ supercell. The calculated results show that AFM-1 is the magnetic ground state for both monolayer Fe$_2$WSe$_4$ and Fe$_2$WS$_4$ [Fig.~\ref{fig2}{f}]. Comparing Figs. \ref{fig2}{(a)} and \ref{fig1}{(a)}, the magnetic and crystal primitive cells are identical, thus Fe$_2$W$X_4$ has no $\left\{ C_2^{\bot}||\tau \right\}$ spin symmetry. Due to the lack of space-inversion symmetry, Fe$_2$W$X_4$ must not have $\left\{ C_2^{\bot}||I \right\}$ spin symmetry. Meanwhile, the Fe$_2$WX$_4$ has $\left\{ C_2^{\bot}||S_{4z} \right\}$ spin symmetry. Therefore, both Fe$_2$WSe$_4$ and Fe$_2$WS$_4$ are the \textit{d}-wave altermagnetic materials.


Then we estimate the N\'eel temperature for monolayer Fe$_2$W$X_4$ by using classical Monte Carlo simulations based on the two-dimensional square lattice Heisenberg model with single-ion anisotropy,
\begin{align}\label{Eq1}
    H=&\sum_{\langle i,j \rangle} J_1 S_i \cdot S_j + \sum_{\langle\langle i,j \rangle\rangle} J_2 S_i \cdot S_j \\\notag
    &+ \sum_{\langle\langle i,j \rangle\rangle} J_2^\prime S_i \cdot S_j + A \sum_i |S_i^z|^2,
\end{align}
where $S_i$ represents the spin on the $i$-site, $J_1$ is the nearest exchange interaction parameter, and $J_2$ and $J_2^\prime$ are the next-nearest exchange interaction parameters with the Fe-Fe channel and Fe-W-Fe channel, respectively [Fig. \ref{fig1} (a)]. $A$ is the single-ion magnetic anisotropy with easy-magnetization axis. These exchange interaction parameters were derived from the four-state method which is a mapping analysis of the four magnetic configurations \cite{4-statesmethod}. Based on the Heisenberg model, we calculated N\'eel temperatures of Fe$_2$WSe$_4$ and Fe$_2$WS$_4$ by classical Monte Carlo simulations \cite{mcTN-2019}, and more details are shown in Appendix \ref{appA:appendix A}. The calculated N\'eel temperatures of Fe$_2$WSe$_4$ and Fe$_2$WS$_4$ are 330K and 500K, respectively. Thus, the altermagnetic order of monolayer Fe$_2$WSe$_4$ and Fe$_2$WS$_4$ remains stable at room temperature.


\begin{figure}[htbp]
\centering
\includegraphics[width=8.5cm]{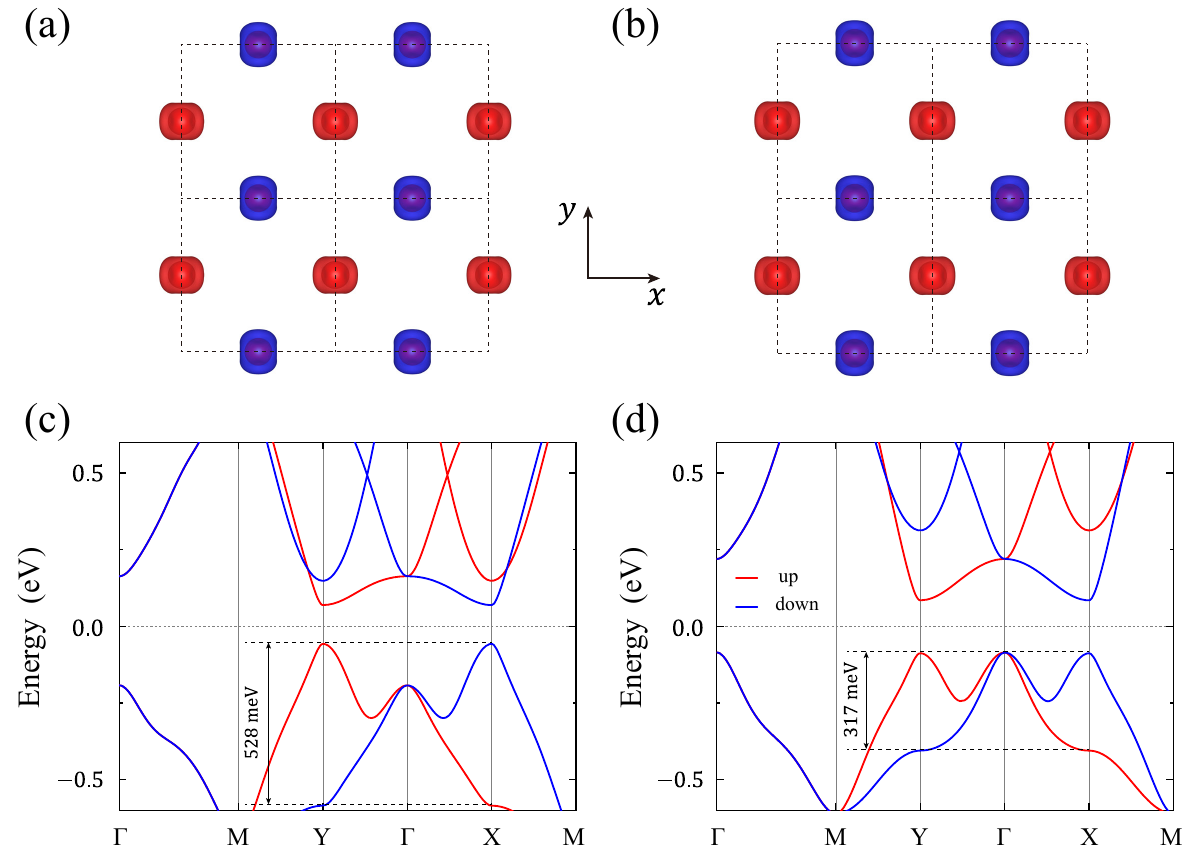}
\caption{(a),(b) Polarization charge densities around the Fe atoms for Fe$_2$WSe$_4$ and Fe$_2$WS$_4$ without SOC (red:  spin-up, blue: spin-down), respectively. (c), (d) are Electronic band structures of Fe$_2$WSe$_4$ and Fe$_2$WS$_4$ along the high-symmetry lines without SOC, respectively.}
\label{fig3}
\end{figure}

After clarifying the magnetic properties, we investigate the electronic properties of monolayer Fe$_2$W$X_4$. From Fig.~\ref{fig3}{(a)}, a spin-up (spin-down) Fe atom has W atoms in the $y$ ($x$) direction but no W atoms in the $x$ ($y$) direction, so the crystal environment of Fe atoms has strong anisotropy, implying that the polarized charge density of Fe atoms may have strong anisotropy. Our calculations show that the polarization charge density of Fe atoms indeed has strong anisotropy [Fig. \ref{fig3}{(a)} and \ref{fig3}{(b)}], which results in the strong anisotropy of Fe-Fe next-neighbor hopping interactions. Moreover, the strong anisotropy of the next-nearest-neighbor hopping interactions can lead to strong spin splitting in altermagnetic materials \cite{BWS}. Therefore, altermagnetic monolayer Fe$_2$W$X_4$ may have large spin splitting. 

Due to $\left\{ C_2^{\bot}||M_{xy} \right\}$ spin symmetry, all bands of Fe$_2$W$X_4$ in the $\Gamma$-$M$ direction are spin degenerate. Considering the characteristics of $d$-wave altermagnetism, all bands at the high-symmetry X and Y points must be spin splitting and have opposite spins due to $\left\{ C_2^{\bot}||M_{xy} \right\}$ spin symmetry. To prove our above analysis, we calculated the band structures of monolayer Fe$_2$W$X_4$ without SOC, which are shown in Figs. \ref{fig3}{(c)} and \ref{fig3}{(d)}. From Fig.~\ref{fig3}{(c) and (d)}, both monolayer Fe$_2$WSe$_4$ and Fe$_2$WS$_4$ are altermagnetic semiconductors with a band gap being 126 meV and 170 meV, respectively. Moreover, both monolayer Fe$_2$WSe$_4$ and Fe$_2$WS$_4$ have large spin splits at certain $k$ points in the BZ, such as high-symmetry X and Y points, to be $528~\rm{meV}$ and $317~\rm{meV}$, respectively. More importantly, both the bottom of the conduction band and the top of the valence band are at the X and Y points and the corresponding bands have opposite spin polarization. This provides a solid foundation for the realization of the crystal valley Hall effect. Due to the 
$\left\{C_2^{\bot}T||C_{2z}T\right\}$ spin symmetry, the Berry curvature is always zero everywhere in the BZ. Then, to realize the crystal valley Hall effect, the SOC effect must be considered for monolayer Fe$_2$W$X_4$.

\begin{figure*}[htbp]
\centering
	\includegraphics[width=17cm]{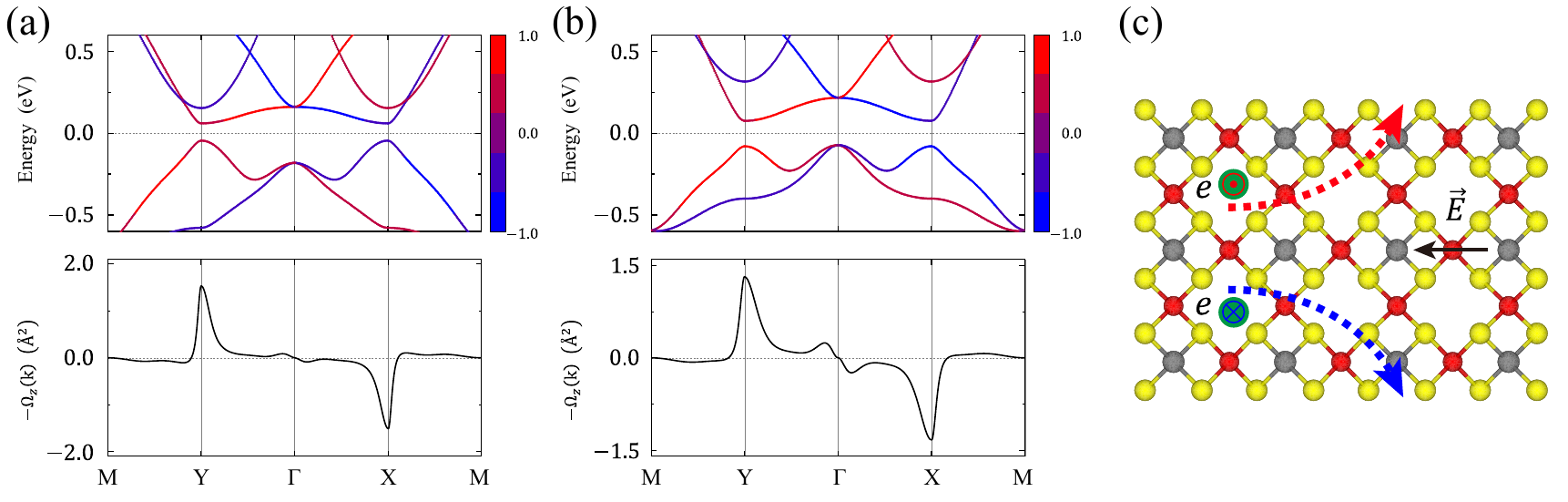}
	\caption{Top panel: Electronic band structure of monolayer (a) Fe$_2$WSe$_4$ and (b) Fe$_2$WS$_4$ with the spin projection $s_z$; bottom panel: the corresponding berry curvature along the high-symmetry line. (c) The schematic diagram of crystal valley Hall effect in monolayer Fe$_2$W$X_4$. $\vec{E}$ and  $e$ represent external in-plane electric field and electrons, respectively. The red and blue arrows (out of plane) represent up and down spins, respectively.}
	\label{fig4}
\end{figure*}

With SOC, the symmetry of monolayer Fe$_2$W$X_4$ changes from spin group to magnetic group. Moreover, the magnetic group symmetry depends on the direction of the easy magnetization axis. To determine the magnetic group symmetry, we calculated the orientation of the easy magnetization axis of monolayer Fe$_2$W$X_4$. The direction of the easy magnetization axis of both monolayer Fe$_2$WSe$_4$ and Fe$_2$WS$_4$ are along the 001 direction. So, both monolayer Fe$_2$WSe$_4$ and Fe$_2$WS$_4$ have 2$S_{4z}T$, $C_{2z}$, $C_{2x}T$, $C_{2y}T$ and 2$M_{xy}$ symmetry. The mirror $M_{xy}$ or $S_{4z}T$ symmetry can protect energy degeneracy and opposite Berry curvatures at the high-symmetry X and Y points [Fig. \ref{fig4}{(a)} and \ref{fig4}{(b)}]. Moreover, the high-symmetry X and Y points still have opposite spin polarization under SOC [Fig. \ref{fig4}{(a)} and \ref{fig4}{(b)}]. These factors will cause monolayer Fe$_2$WX$_4$ to have valley Hall effect under an in-plane electric field [Fig.~\ref{fig4}{(c)}]. Since here a certain crystal symmetry is a necessary condition for realizing the valley Hall effect, we term the valley Hall effect in monolayer Fe$_2$W$X_4$ as the crystal valley Hall effect. Therefore, the valley Hall effect without the $T$ symmetry can be realized in altermagnetic materials.

\begin{figure}[htbp]
	\centering
	\includegraphics[width=8.5cm]{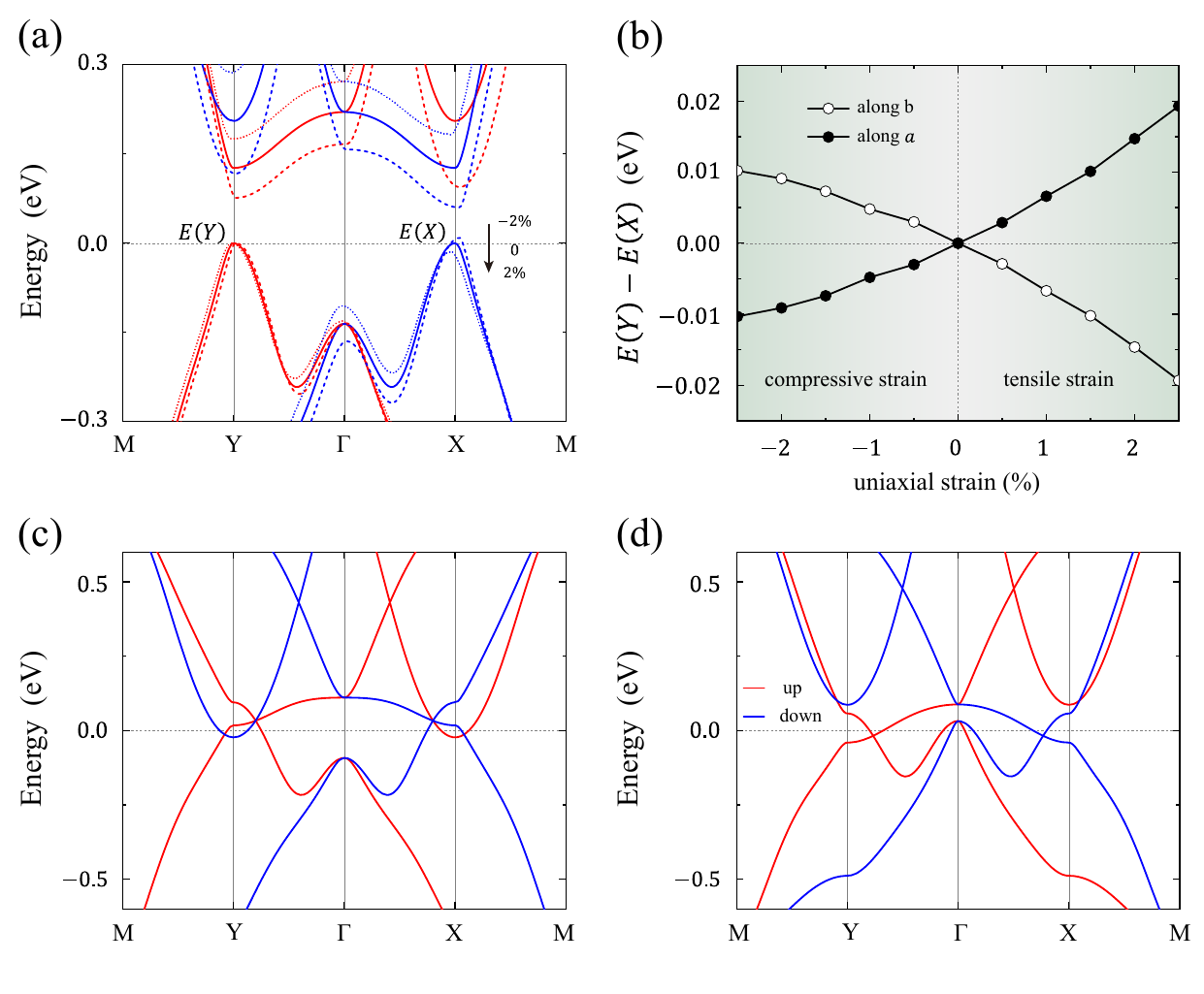}
	\caption{Strain-induced valley polarization and topological phase transition. (a) Band evolution under different uniaxial strains ($-2\%,0,+2\%$) along $a$ direction. (b) Strain-induced valley polarization, which is defined as the energy difference between two valleys $E(Y)-E(X)$. (c), (d) Electronic band structure of Fe$_2$WSe$_4$ and Fe$_2$WS$_4$ without SOC under $3\%$ compression strain along the high-symmetry directions, respectively.}
	\label{fig5}
\end{figure}

Finally, we study the properties of monolayer Fe$_2$W$X_4$ under strain. According to the above analysis, both $\left\{ E||M_{xy} \right\}$ and $\left\{C_2^{\bot} ||C_{4z} \right\}$ spin symmetries can protect the energy band degeneracy at the high-symmetry X and Y points. Applying anisotropic strain to break the $\left\{ C_2^{\bot}||M_{xy} \right\}$ and $\left\{C_2^{\bot} ||C_{4z} \right\}$ spin symmetries will result in band splitting at the high-symmetry X and Y points [Fig.~\ref{fig5}{(a)}]. If we apply compressive stress in the $x$ direction, the spin-down band at the high-symmetry X point will be higher than the spin-up band at the high-symmetry Y point, which corresponds to the spin-down valley polarization [Figs. \ref{fig5}{(a)} and \ref{fig5}{(b)}]. At this point, monolayer Fe$_2$WSe$_4$ is still a semiconductor. When the Fermi level is adjusted to an appropriate position, Fe$_2$WSe$_4$ will have a negative net magnetic moment $M$. Conversely, if we apply a tensile strain in the $x$ direction, monolayer Fe$_2$WSe$_4$ will produce spin-up valley polarization and has a positive net magnetic moment $-M$ by adjusting the Fermi level. Similarly, valley polarization and piezomagnetic effects can be obtained by applying stress along the $y$ direction [Fig.~\ref{fig5}{(b)}]. Our calculations also show that monolayer Fe$_2$WS$_4$ has similar properties. Therefore, both monolayer Fe$_2$WSe$_4$ and Fe$_2$WS$_4$ can realize piezomagnetic effects.

On the other hand, under biaxial compression stress, monolayer Fe$_2$W$X_4$ still has $\left\{ C_2^{\bot}||M_{xy} \right\}$ and $\left\{C_2^{\bot} ||C_{4z} \right\}$ spin symmetries. So the energy bands at high-symmetry X and Y points are still degenerate. Compressive strain can cause band broadening, so monolayer Fe$_2$W$X_4$ may induce band inversion under certain compressive strain due to small band gap. Then, we can realize the topological phase in monolayer Fe$_2$W$X_4$. To prove it, we calculated the electronic band structure of monolayer Fe$_2$W$X_4$ along the high-symmetry direction under $3\%$ compression strain, which is shown in Figs. \ref{fig5}{(c)} and \ref{fig5}{(d)}. From Figs. \ref{fig5}{(c)} and \ref{fig5}{(d)}, both monolayer Fe$_2$WSe$_4$ and Fe$_2$WS$_4$ indeed produce band inversion under $3\%$ compression strain. Moreover, both  monolayer Fe$_2$WSe$_4$ and Fe$_2$WS$_4$ have two pairs of Weyl points protected by $\left\{E ||C_{2x} \right\}$ and $\left\{E ||C_{2y} \right\}$ spin symmetries near the Fermi level and the two pairs of Weyl points have opposite spin polarization [Figs. \ref{fig5}{(c)} and \ref{fig5}{(d)}]. Thus, under compressive strain, both monolayer Fe$_2$WSe$_4$ and Fe$_2$WS$_4$ transform from the semiconductor phase to the bipolarized topological Weyl semimetal phase. Therefore, we can indeed realize topological phases in both monolayer Fe$_2$WSe$_4$ and Fe$_2$WS$_4$ by regulating strain.


In summary, based on spin symmetry analysis and first-principles electronic structure calculations, we predict two two-dimensional altermagnetic semiconductors Fe$_2$WSe$_4$ and Fe$_2$WS$_4$ with magnetic transition temperatures above room temperature. More importantly, when considering SOC, both monolayer Fe$_2$WSe$_4$ and Fe$_2$WS$_4$ can realize crystal valley Hall effect. Since the magnetic transition temperatures of both monolayer Fe$_2$WSe$_4$ and Fe$_2$WS$_4$ exceed room temperature, the crystal valley Hall effect may be observed at room temperature. In addition, under uniaxial strain, both monolayer Fe$_2$WSe$_4$ and Fe$_2$WS$_4$ can achieve piezomagnetic effects. Under biaxial compression stress, both monolayer Fe$_2$WSe$_4$ and Fe$_2$WS$_4$ will transform from the altermagnetic semiconductor phase to the bipolarized Weyl semimetal phase. Therefore, our paper not only proposes unique physical effect, the crystal valley Hall effect, but also provides a good platform to study the crystal valley Hall effect.

\begin{acknowledgments}
This work was financially supported by the National Natural Science Foundation of China (Grant No.12204533, No.12434009, No.62206299 and No.12174443), the National Key R$\&$D Program of China (GrantNo.2024YFA1408601), the Fundamental Research Funds for the Central Universities, and the Research Funds of Renmin University of China (Grant No. 24XNKJ15), the Beijing Natural Science Foundation (Grant No.Z200005) and the Innovation Program for Quantum Science and Technology (2021ZD0302402). Computational resources have been provided by the Physical Laboratory of High Performance Computing at Renmin University of China.
\end{acknowledgments}

\appendix
\begin{appendices}
\section{Some details for calculating the N\'eel temperatures\label{appA:appendix A}}

\begin{figure}[H]
	\centering
	\includegraphics[width=8.5cm]{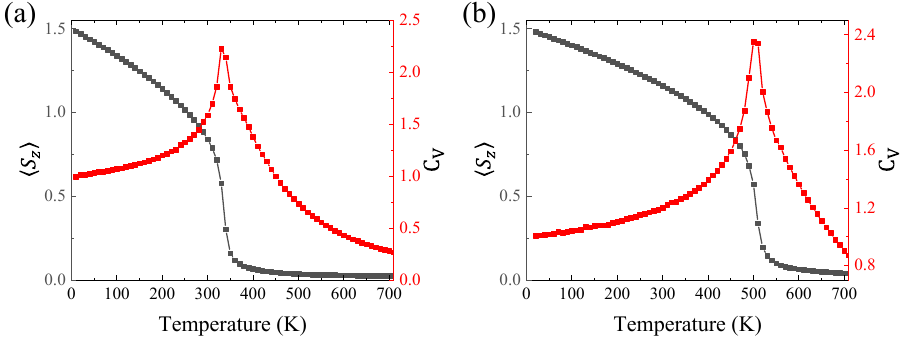}
	\caption{The evolution of $\langle S_z \rangle$ and specific heat capacity with temperature for Fe$_2$WSe$_4$ (a) and Fe$_2$WS$_4$ (b), respectively.}
	\label{fig6}
\end{figure}

We use the four-state method to calculate the exchange interaction parameters in Eq. (1), which are shown in Table \ref{TaB1}. The magnetocrystalline anisotropy energies are 1.000 meV and 0.977 meV for Fe$_2$WSe$_4$ and Fe$_2$WS$_4$, respectively. After performing Monte Carlo simulations by using the MCSOLVER, the resultant N\'eel temperatures can extract from the peak of the specific-heat capacity, which for both Fe$_2$WSe$_4$ and Fe$_2$WS$_4$ are 330 and 500 K, respectively, as shown in Fig. \ref{fig6}.

\begin{table}[H]
\caption{The parameters $J$ in Eq.(\ref{Eq1}) for Fe$_2$W$X_4$ ($X$=Se,S) with $|\mathbf{S}|=3/2$. }
\label{TaB1}
\begin{ruledtabular}
\begin{tabular}{ccc}
\multicolumn{1}{c}{\textrm{Parameters (meV)}}&
 Fe$_2$WSe$_4$ & Fe$_2$WS$_4$ \\
\colrule
$J_1^{xx}$ & 10.987 & 18.157  \\
$J_1^{yy}$ & 10.812 & 18.205 \\
$J_1^{zz}$ & 10.859 & 18.408 \\
$J_2^{xx}$ & $-4.264$ & $-5.533$ \\
$J_2^{yy}$ & $-4.496$ & $-5.968$ \\
$J_2^{zz}$ & $-4.073$ & $-5.977$ \\
${J_2^\prime}^{xx}$ & $-7.025$ & $-7.824$ \\
${J_2^\prime}^{yy}$ & $-6.674$ & $-6.940$ \\
${J_2^\prime}^{zz}$ & $-6.314$ & $-7.471$  
\end{tabular}
\end{ruledtabular}
\end{table}

\section{Molecular dynamics evolutions\label{appB:appendix B}}

\begin{figure}[http]
	\centering
	\includegraphics[width=8.5cm]{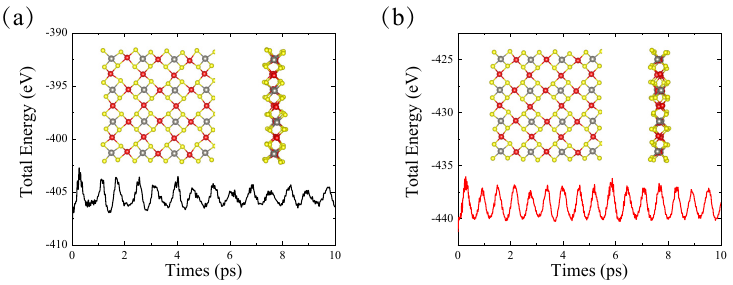}
	\caption{The fluctuation of total energy and the snapshot for the final configuration after molecular dynamics simulation at 300 K for monolayers Fe$_2$WSe$_4$ (a) and Fe$_2$WS$_4$ (b).}
	\label{fig7}
\end{figure}

To further confirm the stability of Fe$_2$WSe$_4$ and Fe$_2$WS$_4$, we calculated their thermodynamic stability by molecular dynamics simulations at 300 K with a 2fs time step over 10 ps, and employing the Nosé-Hoover thermostat for temperature control. Our calculated results show that both Fe$_2$WSe$_4$ and Fe$_2$WS$_4$ are in thermodynamic stability (Fig. \ref{fig7}).

\end{appendices}

\nocite{*}

\bibliography{Reference}

\end{document}